\documentclass[sigconf,screen, natbib=true]{acmart}
\usepackage{graphicx}
\usepackage{subcaption}
\usepackage{algorithm}
\usepackage{algpseudocode}
\usepackage{enumitem}
\usepackage{multirow}
\usepackage{tabularx}
\usepackage{threeparttable}
\usepackage{color, xspace}
\usepackage{appendix}
\usepackage{balance}

\AtBeginDocument{%
	\providecommand\BibTeX{{%
			\normalfont B\kern-0.5em{\scshape i\kern-0.25em b}\kern-0.8em\TeX}}}
\copyrightyear{2025}
\acmYear{2025}
\setcopyright{acmlicensed}\acmConference[KDD '25]{Proceedings of the 31st ACM
SIGKDD Conference on Knowledge Discovery and Data Mining V.1}{August 3--7,
2025}{Toronto, ON, Canada}
\acmBooktitle{Proceedings of the 31st ACM SIGKDD Conference on Knowledge
Discovery and Data Mining V.1 (KDD '25), August 3--7, 2025, Toronto, ON, Canada}
\acmDOI{10.1145/3690624.3709304}
\acmISBN{979-8-4007-1245-6/25/08}

\begin{document}
\title[GLINT-RU: Gated Lightweight Intelligent Recurrent Units for Sequential Recommender Systems]{GLINT-RU: Gated Lightweight Intelligent Recurrent Units for Sequential Recommender Systems}

\author{Sheng Zhang}
\authornote{Equal contribution.}
\affiliation{%
  \institution{City University of Hong Kong and}
  \streetaddress{}
  \city{}
  \state{}
  \country{}
}
\affiliation{%
  \institution{High Magnetic Field Laboratory, Chinese Academy of Sciences}
  \streetaddress{}
  \city{Hefei and Hong Kong}
  \state{}
  \country{China}
}
\email{szhang844-c@my.cityu.edu.hk}

\author{Maolin Wang}
\authornotemark[1] 
\affiliation{%
  \institution{City University of Hong Kong}
  \streetaddress{}
  \city{Hong Kong}
  \state{}
  \country{China}
}
\email{morin.wang@my.cityu.edu.hk}

\author{Wanyu Wang}
\authornote{Corresponding author.}
\affiliation{%
  \institution{City University of Hong Kong}
  \streetaddress{}
  \city{Hong Kong}
  \state{}
  \country{China}
  \postcode{}
}
\email{wanyu.wang@my.cityu.edu.hk}

\author{Jingtong Gao}
\affiliation{%
  \institution{City University of Hong Kong}
  \streetaddress{}
  \city{Hong Kong}
  \state{}
  \country{China}
  \postcode{}
}
\email{jingtong.gao@my.cityu.edu.hk}

\author{Xiangyu Zhao}
\affiliation{%
  \institution{City University of Hong Kong}
  \streetaddress{}
  \city{Hong Kong}
  \state{}
  \country{China}
  \postcode{}
}
\email{xianzhao@cityu.edu.hk}

\author{Yu Yang}
\affiliation{%
  \institution{City University of Hong Kong}
  \streetaddress{}
  \city{Hong Kong}
  \state{}
  \country{China}
  \postcode{}
}
\email{yu.yang@cityu.edu.hk}

\author{Xuetao Wei}
\affiliation{%
  \institution{Southern University of \\Science and Technology}
  \streetaddress{}
  \city{Shenzhen}
  \state{}
  \country{China}
  \postcode{}
}
\email{weixt@sustech.edu.cn}

\author{Zitao Liu}
\affiliation{%
  \institution{Jinan University}
  \streetaddress{}
  \city{Guangzhou}
  \state{}
  \country{China}
  \postcode{}
}
\email{liuzitao@jnu.edu.cn}

\author{Tong Xu}
\affiliation{%
  \institution{University of Science \\and Technology of China}
  \streetaddress{}
  \city{Hefei}
  \state{}
  \country{China}
  \postcode{}
}
\email{tongxu@ustc.edu.cn}

\renewcommand{\shortauthors}{Sheng Zhang, et al.}
\begin{abstract}

Transformer-based models have gained significant traction in sequential recommender systems (SRSs) for their ability to capture user-item interactions effectively. However, these models often suffer from high computational costs and slow inference. Meanwhile, existing efficient SRS approaches struggle to embed high-quality semantic and positional information into latent representations.
To tackle these challenges, this paper introduces \textbf{GLINT-RU}, a lightweight and efficient SRS leveraging a single-layer dense selective Gated Recurrent Units (GRU) module to accelerate inference. By incorporating a dense selective gate, GLINT-RU adaptively captures temporal dependencies and fine-grained positional information, generating high-quality latent representations. Additionally, a parallel mixing block infuses fine-grained positional features into user-item interactions, enhancing both recommendation quality and efficiency.
Extensive experiments on three datasets demonstrate that GLINT-RU achieves superior prediction accuracy and inference speed, outperforming baselines based on RNNs, Transformers, MLPs, and SSMs. These results establish GLINT-RU as a powerful and efficient solution for SRSs. The implementation code is publicly available for reproducibility.~\footnote{https://github.com/szhang-cityu/GLINT-RU}.

\end{abstract}




\maketitle
\section{Introduction}
In this era of data exploding, sequential recommender systems (SRSs) ~\cite{DL4, LRU, Linrec, GRU4Rec, Kang01, mamba4rec, mlp4rec, liu2024bidirectional, zhao2023embedding} have gained much attention in capturing users' preferences within a large amount of sequential interaction data. GRU4Rec ~\cite{GRU4Rec} as one of the earliest session-based recommendation models, employs stacked Gated Recurrent Units (GRU) for item-to-item recommendations. However, the RNN-based methods~\cite{GRU4Rec, narm} are fading from the recommendation realm due to their relatively lower accuracy. In recent years, transformer-based SRSs ~\cite{Kang01, bert4rec} have become increasingly popular for the powerful multi-head attention mechanism ~\cite{zhao2023user,vaswani01}. They exhibit remarkable ability in capturing sequential interactions and delivering accurate predictions~\cite{strec}. However, despite their effectiveness, current transformer-based models suffer from substantial computational demands and extended inference time, which is caused by the dot product operation in the attention mechanism ~\cite{Linformer, Linrec, zhou2023opportunities}. 

To tackle the issue of the high computational cost of transformer-based SRSs, linear attention mechanisms are applied to reduce the computational complexity. For example, LinRec ~\cite{Linrec} changes the order of dot product between query and key matrices by designing a special mapping, dramatically reducing the inference time. LightSAN ~\cite{lightsan} projects historical interactions into interest representations with shorter lengths, thereby reducing the computational complexity of transformers to a linear scale. MLP-based frameworks~\cite{mlp4rec, FMLP, MLM4Rec} can also achieve fast inference speed and high performance by reshaping input sequence tensors~\cite{smlp4rec}. SSM-based models ~\cite{mamba4rec, MambaRec} outperforms the existing attention mechanism by utilizing the efficient selective SSM ~\cite{gu2023mamba}, within which a structured state tensor is used to address long-range dependencies.

However, to achieve high performance, the transformer-based SRSs, even with linear attention mechanism, require deeply stacked transformers, which decreases the model efficiency ~\cite{easrec}. 
MLP-based SRSs with sequence mixing layers might suffer from extended inference time when dealing with long sequences. In addition, MLP-based models struggle to capture fine-grained positional dependencies. SSM-based models~\cite{mamba4rec} may struggle to model effective semantic features into latent representations in long/short-term recommendation scenarios, as they might have difficulty learning important interactions.
In this paper, we aim to further improve the efficiency and the accuracy of efficient models in various scenarios.

To further reduce resource consumption, accelerate inference speed, and enhance the model performance, we propose a novel efficient SRS framework named \textbf{G}ated \textbf{L}ightweight \textbf{I}ntellige\textbf{NT} \textbf{R}ecurrent \textbf{U}nits (GLINT-RU). Considering that stacking hybrid architectures may lead to a deeper model, which will cause a significant decrease in inference speed. and the traditional GRU module lacks the ability to adaptively adjust the output and filter out unimportant interactions. To tackle these challenges, we employ various gate mechanisms in appropriate positions to fully perceive the data environment and filter information automatically ~\cite{easrec, PEPNet, griffin}. We introduce an expert mixing block that captures the item dependencies via GRU and utilizes linear attention to capture the global interaction information between users and items~\cite{moe, moesurvey}. This strategy not only improves the inference speed due to the linear computational complexity of paralleled GRU and linear attention mechanism but also enhances the context information. Additionally, we implement a dense selective GRU, which selects the output of the GRU adaptively and considers the connections among adjacent items. It leverages the gate mechanism to select crossed channels and extracts short-duration patterns to refine the model's understanding of user behavior dynamics. Moreover, a gated MLP block is utilized to select the outputs of the expert mixing block, deeply filtering the information based on the data environment.

We summarize the major contributions of our work as follows:
\vspace{-\topsep}
\begin{itemize}[leftmargin=*]
    \item In this paper, we introduce GLINT-RU, a novel and lightweight SRS that achieves remarkable inference speed only requiring a single layer. It is an advanced model that captures complex semantic features and fine-grained positional representations using an expert mixing mechanism, which substantially improves the performance of GLINT-RU over existing efficient SRS models.
    \item We introduce a dense selective GRU module, which not only incorporates connections between adjacent items but also empowers the model with the capability to selectively learn long-term dependencies. The integration of this advanced GRU module into the model markedly elevates its performance, establishing a new standard for efficient recommender systems.
    \item We conduct extensive experiments to verify the efficiency and effectiveness of GLINT-RU on various datasets and parameter settings. Our GLINT-RU improves the training and inference speed significantly and stably exhibits high performance.
\end{itemize}
\vspace{-\topsep}

\section{Preliminaries}
In this section we will briefly introduce our recommendation task, and then introduce the basic efficient GRU and linear attention modules used in our framework.
\subsection{Problem Statement}
For a sequential recommendation task, we have a set of users $\mathcal{U} = \{u_1, u_2, \dots, u_{\vert \mathcal{U} \vert}\}$ who have historical interactions with a set of items $\mathcal{V}=\{v_1, v_2,\dots, v_{\vert \mathcal{V} \vert}\}$. Among these users, the $i$-th user has a preferred item sequence denoted as $s_i=[v_1^{(i)}, v_2^{(i)}, \dots$, $v_{n_i}^{(i)}]$, where $n_i$ is the length of the item list that the $i$-th user interacts with. Our goal is to design an efficient framework and predict the rating score of the next item based on the historical interactions .
\subsection{Gated Recurrent Units}
As an essential part of GLINT-RU, the GRU ~\cite{GRU} module contributes to the recommendation task by capturing the dependencies among the items and dynamically adjust its memory content
~\cite{GRU2, GRU3}. The update mechanism of a GRU cell is formulated as follows:
\begin{equation}
    \begin{aligned}
        \boldsymbol{z}_t &= \sigma(\boldsymbol{W}_z \cdot [\boldsymbol{h}_{t-1}, \boldsymbol{x}_t] + \boldsymbol{b}_z),\\
        \boldsymbol{r}_t &= \sigma(\boldsymbol{W}_r \cdot [\boldsymbol{h}_{t-1}, \boldsymbol{x}_t] + \boldsymbol{b}_r),\\
        \tilde{\boldsymbol{h}}_t &= \tanh(\boldsymbol{W} \cdot [\boldsymbol{r}_t * \boldsymbol{h}_{t-1}, \boldsymbol{x}_t] + \boldsymbol{b}),\\
        \boldsymbol{h}_t &= \boldsymbol{z}_t * \boldsymbol{h}_{t-1} + (1 - \boldsymbol{z}_t) * \tilde{\boldsymbol{h}}_t,
    \end{aligned}
    \label{grucell}
\end{equation}
where $\sigma(\cdot)$ is the sigmoid activation function, $\boldsymbol{x}_t$ is the input of GRU module in $t$-th time step, $\boldsymbol{h}_t$ represents the $t$-th hidden states, $\boldsymbol{z}_t$ and $\boldsymbol{r}_t$ are the update gate and the reset gate, respectively. $\boldsymbol{b}_z, \boldsymbol{b}_r, \boldsymbol{b}$ are bias, $\boldsymbol{W}_z, \boldsymbol{W}_r, \boldsymbol{W}$ are trainable weight matrices. As is shown in Eq.(\ref{grucell}), GRU uses the update gate to control the retained information volume from previous hidden states in the current time step, while the reset gate controls the information that should be forgotten. 

The GRU (Gated Recurrent Unit) module, equipped with update and reset gates in sequential GRU Cells, is adept at capturing the relationships among the items throughout the sequence while maintaining a relatively low computational complexity. However, the sequential information in GRU cannot be interacted with, and each hidden state is primarily encoded from preceding elements, which restricts the representational capacity of GRU-based recommendation models, particularly in capturing complex item dependencies across the entire sequence.
\subsection{Linear Attention Mechanism}
The attention mechanism as a powerful core of the transformer structure, exhibits performance in learning sequence interactions in recommendation tasks. However, the high computational cost of the dot product between query matrix $\boldsymbol{Q}$ and key matrix $\boldsymbol{K}$ substantially lower the inference speed of transformer-based SRSs especially when the sequence length $N$ is much larger than hidden size $d$~\cite{EA, Linformer}. To tackle this issue, the linear attention mechanism ~\cite{Linrec} designs a special mapping function to change the order of the dot product and reduce the computational complexity to $\mathcal{O}(Nd^2)$. The linear attention mechanism can be written as:
\begin{equation}
\begin{aligned}
    \boldsymbol{A}'(\boldsymbol{Q,K,V}) = \mathcal{X}_1\left(elu(\boldsymbol{Q})\right) \left(\mathcal{X}_2\left(elu(\boldsymbol{K})\right)^{\mathrm{T}} \boldsymbol{V}\right),
\end{aligned}
\end{equation}
where $\mathcal{X}_1$ and $\mathcal{X}_2$ are row-wise and column-wise $L_2$ normalization mappings, respectively, $\boldsymbol{Q,K,V}$ are learnable query, key and value matrices and $\boldsymbol{A}'$ is the output attention score. This approach mitigates the issue that the softmax layer concentrates on the scores of merely a few positions, enlarging the information capacity of the attention mechanism ~\cite{Linrec}. By implementing linear attention, our GLINT-RU framework is capable of learning interactions between items in long sequences.
\section{METHODOLOGY}
	\begin{figure*}[t]
        \centering
        \includegraphics[width=0.93\textwidth]{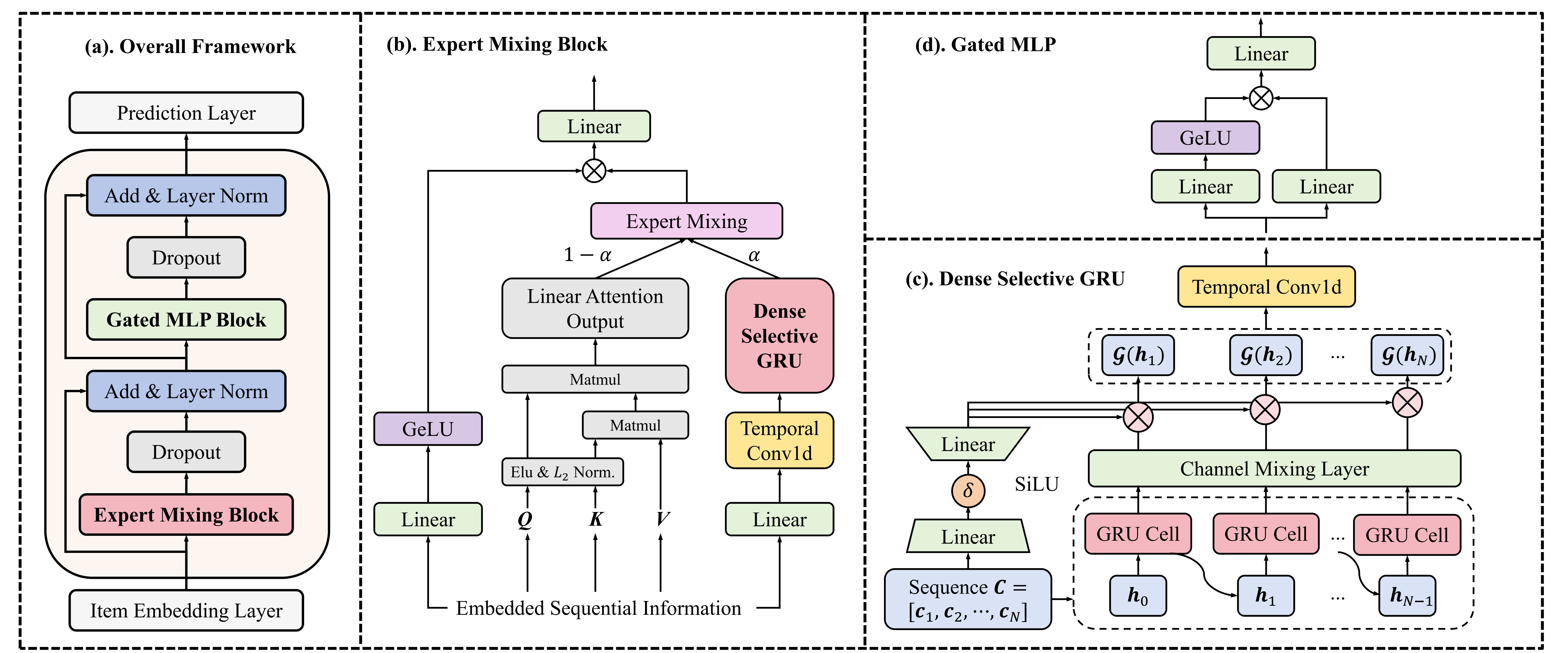}
        \caption{\textbf{(a). Framework of proposed GLINT-RU. (b). Expert mixing block employs paralleled attention and GRU to effectively learn semantic features and fine-grained positional information. (c). Dense Selective GRU as the core part of the framework, deeply selects and aggregates the hidden states. (d). Gated MLP block is utilized to deeply filter the feed forward network.}}
    \label{fig:network}
    \vspace{-3mm}
\end{figure*}
In this section, we will introduce the overall framework and its components that effectively capture semantic features and positional information, followed by the complexity analysis of GLINT-RU.
\subsection{Framework Overview}
Many existing recommendation frameworks depend on transformer structure ~\cite{bert4rec, Kang01, Linrec}, which incurs substantial computational overhead and low inference speed. Restricted by the large computational complexity of stacked transformers, linear attention-based models have approached a plateau in terms of minimizing inference time and resource consumption. Uniquely, we propose an advanced recommendation framework that integrates the linear attention mechanism and efficient dense selective GRU module, which further reduces the computational cost compared with stacked linear transformers and SSM-based models. Additionally, this dense selective GRU module also enables our framework to understand both semantic features and dependencies from item sequences, and substantially reduce the computational cost and inference time. 

Figure \ref{fig:network}.(a) shows the structure of our GLINT-RU.
As is shown in Figure \ref{fig:network}.(b)-(d), GLINT-RU integrates an \textbf{expert mixing block} for mixing sequential information from the \textbf{dense selective GRU} expert and the linear attention expert, and a \textbf{gated MLP block} for further learning and filtering complex user behaviors. 

In the \textbf{expert mixing block}, the \textbf{dense selective GRU} module is employed to capture the long/short-term item-wise dependencies, and selectively learn the sequential information. In addition, the linear attention expert is responsible for modeling item interactions from the user. By combining these two powerful expert modules, our GLINT-RU is capable of adaptively learning both temporal and semantic item features from the sequence, which performs fine-grained modeling of complex user behaviors.

After the user-item interactions are selectively learned by mixing block, the item scores are conveyed to the \textbf{gated MLP block}, where the information is filtered according to the data environment. The framework employs various gates in appropriate positions to deeply filter information, improving the model's flexibility and the ability to perceive and select information.

\subsection{Item Embedding Layer}
For sequential recommendation tasks, information on items interacted by users should be encoded to tensors through the embedding layer ~\cite{Embedding}. We denote the length of input user-item interactions as $N$, and embedding size as $d$. For a interaction sequence $s_i=[v_1, v_2, \dots, v_n, \dots v_{n_i}]$, the $n$-th item $v_n\in \mathbb{R}^{D_n}$ can be projected into the representation $\boldsymbol{e}_n$ by the following formulation:
\begin{equation}
    \boldsymbol{e}_n = \boldsymbol{W}_n v_n, 
\end{equation}
where $\boldsymbol{W}_n \in \mathbb{R}^{d\times D_n}$ is trainable weighted matrix. The embedding layer outputs the encoded item sequence in a tensor:
\begin{equation}
    \boldsymbol{E} = [\boldsymbol{e}_1, \boldsymbol{e}_2, \cdots, \boldsymbol{e}_N]^{\mathrm{T}}.
\end{equation}

In traditional transformer-based models, positional embeddings are typically necessary because the attention mechanism lacks the inherent capability to encode temporal information ~\cite{Kang01}. Uniquely, in this paper we employ the GRU module to model temporal item dependencies, which generates fine-grained representations with positional information for the items. Therefore, we decide not to add the positional embedding layer into the framework.
\subsection{Dense Selective GRU}
Existing GRU cell learns sequential data by conveying information from preceding cells. Although this mechanism has the superiority of capturing the long-term dependencies in the sequence, it predominantly focuses on information from previous items, while potentially overlooking the valuable context information provided by adjacent items, which are often closely related in real-world applications.
To address these challenges, we introduce dense selective GRU shown in Figure \ref{fig:network}.(b) as the core component of GLINT-TU. This innovation extracts local temporal features and generates fine-grained positional information using the update mechanism in GRU module.
By implementing dense selective GRU, the computational complexity can be further reduced, and the recommendation accuracy of the GRU-based framework can be substantially improved.
\subsubsection{\textbf{Dense GRU module}}
Therefore, to enable each GRU cell to aggregate local temporal features of user behavior, we introduce a temporal convolution layer , where adjacent item information is adaptively fused before being fed into the GRU module:
\begin{equation}
    \boldsymbol{C} = \mathrm{TemporalConv1d}(\boldsymbol{\boldsymbol{X}\boldsymbol{W}_0 + \boldsymbol{b}_0})
\end{equation}
where $\boldsymbol{X} = [x_1, x_2, \dots, \boldsymbol{x}_N]^{\mathrm{T}}$ is the input tensor with $d$ feature dimensions, and $\boldsymbol{C}=[\boldsymbol{c}_1, \boldsymbol{c}_2, \dots, \boldsymbol{c}_N]^{\mathrm{T}}$ is the output of the convolution operation $\mathrm{TemporalConv1d(\cdot)}$ with $N$ steps. $\boldsymbol{W}_0$ and $\boldsymbol{b}_0$ are weight matrix and bias, respectively. The size of the convolution kernel is set as $k$. According to Eq.(\ref{grucell}), the output of the $\mathrm{GRUCell}(\cdot)$  can be divided into a latent item representation $\hat{\boldsymbol{h}}_n$ and a fine-grained positional representation $\boldsymbol{p}_n$ learned by historical hidden states:
\begin{equation}
\begin{aligned}
    \boldsymbol{h}_n &= \mathrm{GRUCell}(\boldsymbol{c}_n, \boldsymbol{h}_{n-1}) = \hat{\boldsymbol{h}}_n + \boldsymbol{p}_n \\
    \hat{\boldsymbol{h}}_n &= (1-\boldsymbol{z}_t)\tilde{\boldsymbol{h}}_n + \prod\limits_{i=1}^n \boldsymbol{z}_i \boldsymbol{h}_0, \ \ \boldsymbol{p}_n = \sum\limits_{k=1}^{n-2} \prod\limits_{i=k+1}^{n} \boldsymbol{z}_i(1-\boldsymbol{z}_k)\boldsymbol{h}_k
\end{aligned}
\end{equation}
where $\boldsymbol{z}_k$ is the reset gate at the $k$-th time step. It is noteworthy that each positional representation $\boldsymbol{p}_n$ is generated by aggregated historical hidden states with varied update intensities.
Then we project the hidden states into a latent space using the channel crossing layer, which can be written as:
\begin{equation}
    \Phi(\boldsymbol{H}) = \boldsymbol{H}\boldsymbol{W}_{H}+\boldsymbol{b}_H,
\end{equation}
where $\boldsymbol{H} = [\boldsymbol{h}_1, \boldsymbol{h}_2, \dots, \boldsymbol{h}_N]^{\mathrm{T}}$ is the output of GRU, $\boldsymbol{W}_H$ is the learnable weight matrix and $b_H$ are bias. Although each input state $\boldsymbol{c}_t$ incorporates information from both preceding and subsequent items, each output hidden state is still determined by the hidden state of the preceding time step. Therefore, to capture the context information of the output sequential hidden states, we implement a temporal convolution on the crossed features. This convolution layer extracts local temporal features to understand user behavior dynamics, and enhance the predictive accuracy of our model:
\begin{equation}
    \boldsymbol{Y} =  \mathrm{TemporalConv1d}(\mathcal{G})
\end{equation}
where $\mathcal{G}$ is the Selective Gate function, and $\boldsymbol{Y} = [\boldsymbol{y}_1, \boldsymbol{y}_2, \dots, \boldsymbol{y}_N]^{\mathrm{T}}$ is the output matrix from the dense selective GRU module. The two convolution layers together with GRU cells improve the density of the sequential information, enabling each hidden state to be learned from behaviors of more input time steps.
\subsubsection{\textbf{Selective Gate}}
To filter the hidden information of the GRU module, we design a selective gate where outputs of the feature crossing layer are selected based on the input state of the GRU. The selective gate weights are generated by two tiny linear layers with SiLU activation function ~\cite{gate}, and we use them to select useful hidden states and filter information:
\begin{equation}
    \begin{aligned}
        \boldsymbol{\delta}_1(\boldsymbol{C}) &= \boldsymbol{C}\boldsymbol{W}_{\delta}^{(1)}+b_{\delta}^{(1)},\\
        \mathcal{G}(\boldsymbol{H}) = \boldsymbol{\Omega}(\boldsymbol{\delta}_1(\boldsymbol{C}), \boldsymbol{H})
        &= (\mathrm{SiLU}(\boldsymbol{\delta}_1(\boldsymbol{C}))\boldsymbol{W}_{\Omega}^{(1)}+\boldsymbol{b}_{\Omega}^{(1)})\otimes \Phi(\boldsymbol{H}),
    \end{aligned}
\end{equation}
where $\boldsymbol{C}=[\boldsymbol{c}_1, \boldsymbol{c}_2, \dots, \boldsymbol{c}_N]^{\mathrm{T}}$ also serves as the input of the GRU module, $\boldsymbol{W}_{\delta}^{(1)}$ , $\boldsymbol{W}_{\Omega}^{(1)}$ are weight matrices, $\boldsymbol{b}_{\delta}^{(1)},\boldsymbol{b}_{\Omega}^{(1)}$ are bias. With this gate mechanism, our GRU-based model could become more flexible with the ability to perceive the data environment.

\subsection{Expert Mixing Block}
\begin{figure}[t]
        \centering
        \includegraphics[width=0.85\linewidth]{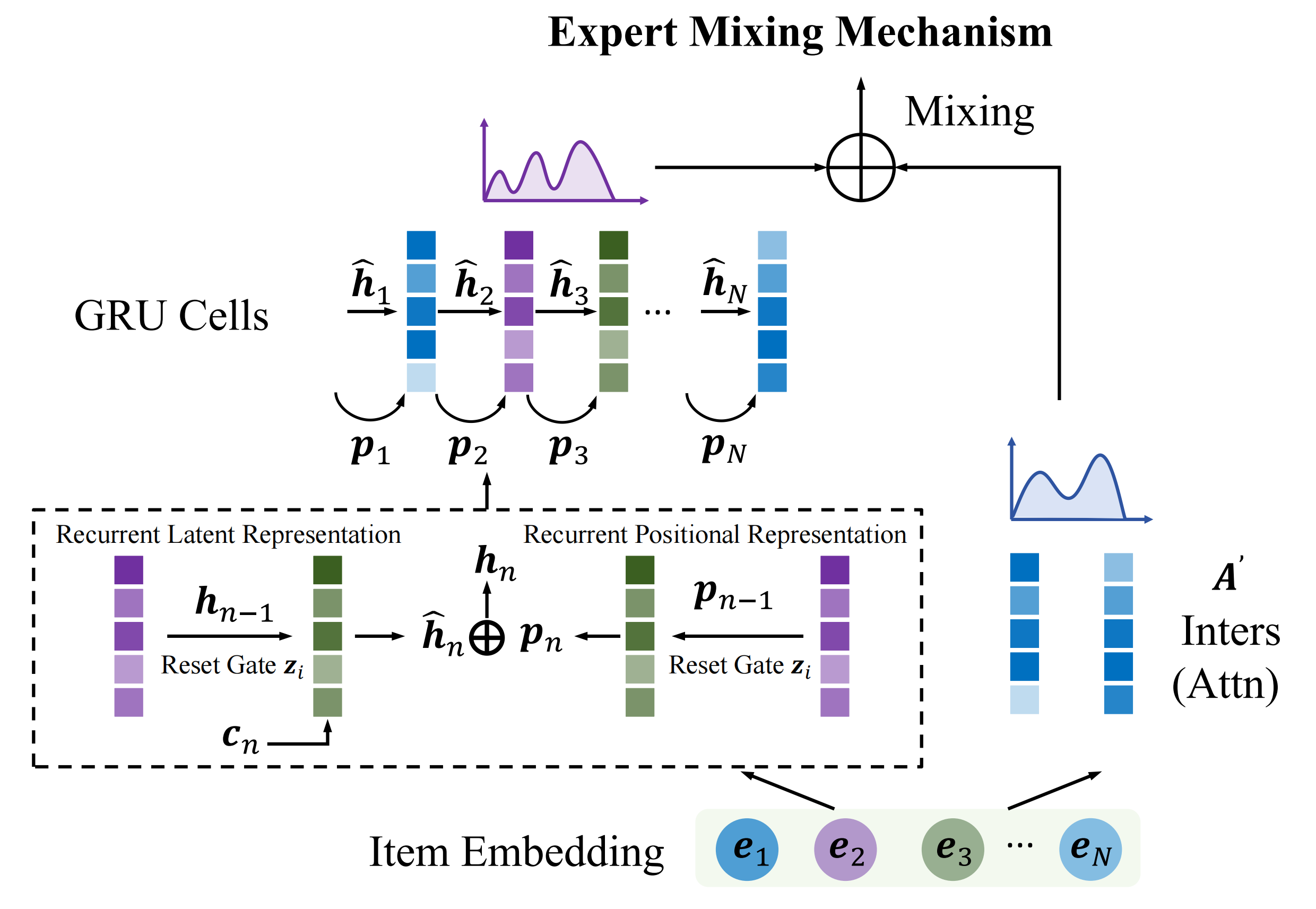}
        \caption{\textbf{General process of expert mixing mechanism. GRU captures long-term dependencies with recurrent latent and fine-grained positional representations. Attention layer learns semantic features fro important item interactions.}}
    \label{fig:mixing}
    \vspace{-5mm}
\end{figure}

Existing efficient SRSs lack the ability to exploit semantic features, fine-grained positional information, and dependencies simultaneously. As is shown in Figure \ref{fig:network}.(c), by introducing the linear attention mechanism and mixing these two powerful experts, the computational complexity will be substantially reduced compared with transformer-based models, and more high-quality item representations can be generated.
Moreover, the two employed experts are parallel in our framework, which further improves the model's efficiency. The expert mixing mechanism is shown in Figure \ref{fig:mixing}.

In real applications, the conditions of the data vary a lot. The GRU is naturally suited for sequential data, demonstrating effectiveness in sequential recommendation tasks that exhibit strong temporal dependencies, while attention focuses on relevant items in the sequence dynamically. To adapt to complex data conditions, we attribute appropriate weights to the two experts by a mixing gate:
\begin{equation}
\label{mixing}
    \begin{aligned}
        \boldsymbol{M} &= \alpha_1^{(t)} \boldsymbol{A}'(\boldsymbol{Q}, \boldsymbol{K}, \boldsymbol{V}) + \alpha_2^{(t)}\boldsymbol{Y}, \\
        \alpha_i^{(t)} &= \mathrm{softmax}(\alpha_i^{(t-1)}) = \frac{\alpha_i^{(t-1)}}{e^{\alpha_1^{(t-1)}}+e^{\alpha_2^{(t-1)}}}, \ \ i=1,2,
    \end{aligned}
\end{equation}
where $\alpha_1^{(t)}, \alpha_2^{(t)}$ are trainable mixing parameters at $t$-th training iteration. Then we filter this output by introducing another data-aware gate which selects the outputs based on the original input data batch:
\begin{equation}
    \begin{aligned}
        \boldsymbol{\delta}_2(\boldsymbol{X}) &= \boldsymbol{X}\boldsymbol{W}_{\delta}^{(2)}+b_{\delta}^{(2)},\\
        \boldsymbol{Z} = \boldsymbol{\Omega}_2(\boldsymbol{\delta}_2(\boldsymbol{X}), \boldsymbol{M}) &= (\mathrm{GeLU}(\boldsymbol{\delta}_2(\boldsymbol{X})))\otimes \boldsymbol{M},
    \end{aligned}
\end{equation}
where $\boldsymbol{X}$ is the input of expert mixing block, $\boldsymbol{W}_{\delta}^{(2)}$ is the weight matrix of linear layer, $\boldsymbol{b}_{\delta}^{(2)}$ are bias.

\subsection{Gated MLP Block}
Most existing efficient SRSs, utilize two-layer feed forward networks to capture the nonlinear relationships among features before giving predictions. To further enhance the performance of the model and augment useful features from the expert mixing block, we introduce the gated MLP block shown in Figure \ref{fig:network}.(d), which employs a gate mechanism again to deeply filter the information and generate item representations for predictions.
\begin{equation}
    \begin{aligned}
        \boldsymbol{\delta}_3(\boldsymbol{Z}) &= \boldsymbol{Z}\boldsymbol{W}_{\delta}^{(3)}+b_{\delta}^{(3)},\\
        \boldsymbol{P} = \boldsymbol{\Omega}_2(\boldsymbol{\delta}_3(\boldsymbol{Z}), \boldsymbol{Z}) &= (\mathrm{GeLU}(\boldsymbol{\delta}_3(\boldsymbol{Z})))\otimes (\boldsymbol{Z}\boldsymbol{W}+\boldsymbol{b}),\\
        \boldsymbol{R} &= \boldsymbol{P}\boldsymbol{W}_o + \boldsymbol{b}_o
    \end{aligned}
\end{equation}
where $\boldsymbol{Z}$ is the output of expert mixing block, $\boldsymbol{P}$ denotes the output of gated linear layer, $\boldsymbol{R}$ represents the item representation, and $\boldsymbol{W}_{\delta}^{(3)}, \boldsymbol{W}_o, \boldsymbol{W}$ are weight matrices, $\boldsymbol{b}_{\delta}^{(3)}, \boldsymbol{b}_o, \boldsymbol{b}$ are bias. The recommendation scores are generated by item representations and embeddings, followed by the prediction score $\hat{y}_i$ of the $i$-th item:
\begin{equation}
\label{score}
    \begin{aligned}
        \hat{y}_i = \mathrm{softmax}(\boldsymbol{R}_i(\boldsymbol{e}_i)^{\mathrm{T}}),
    \end{aligned}
\end{equation}
where $\boldsymbol{R}_i$ is the representation of $i$-th item. Loss function, model training methods and the algorithm are displayed in \textbf{Appendix \ref{training}}.

\subsection{Complexity Analysis}
\label{complexity}
In this subsection, we will explain why GLINT-RU has inherent superiority over other popular SRS models in model efficiency.

Given that the sequence length is $N$, the embedding size is $d$ and the kernel size for GLINT-RU is $k$, the time complexity of GLINT-RU is $\mathcal{O}((2k+12)Nd^2)$. The complexity is calculated throughout the network, from the embedding layer to the prediction layer. 
\textbf{Discussion.}
Our GLINT-RU shows significantly low and linear time complexity, as GLINT-RU is a highly paralleled mixed network with only one layer to achieve high performance. Our framework utilizes paralleled expert networks and employs an efficient GRU module to capture long-term dependencies. GLINT-RU is more efficient than other models in the following aspects: 
\begin{itemize}[leftmargin=*]
    \item \textbf{GLINT-RU v.s. Transformer-based SRSs}: Firstly, traditional transformer-based model ~\cite{Kang01} suffers from large computational complexity, especially when the sequence length $N$ is large. Linear attention mechanism ~\cite{Linrec} can be applied to substantially reduce the computational cost. However, they still require stacked transformer structures to achieve high performance, while GLINT-RU achieves outstanding performance with only one layer. 
    \item \textbf{GLINT-RU v.s. MLP-based SRSs}: Secondly, The inference speed of the MLP-based models~\cite{smlp4rec} may decrease when faced with long sequence length, as the sequence mixing layer has quadratic time complexity. In contrast, the paralleled expert module of GLINT-RU is more suitable for processing long sequential data.
    \item \textbf{GLINT-RU v.s. SSM-based SRSs}: Thirdly, state-space models~\cite{gu2023mamba} may require complex matrix operations and recursive calculations, which may be difficult to parallelize efficiently in practical calculations. In contrast, GLINT-RU's paralleled expert module has a simpler update mechanism and higher efficiency.
\end{itemize}


\section{Experiments}
    In this section, we conduct extensive experiments to show the effectiveness and efficiency of our GLINT-RU Framework. After we introduce our implementation details, the experiment results will be analyzed in detail. The experiments in this section are set to answer the following research questions:
    
    \begin{itemize}[leftmargin=*]
        \item \textbf{RQ1:} How does GLINT-RU framework perform compared with other state-of-the-art SRS baseline models?
        \item \textbf{RQ2:} To what extent does GLINT-RU improve model efficiency compared with other state-of-the-art SRS frameworks?
        \item \textbf{RQ3:} How do the dense selective GRU, the linear attention mechanism, and the gated MLP contribute to GLINT-RU?
        \item \textbf{RQ4:} How does the hyperparameter setting affect the performance of GLINT-RU?
    \end{itemize}
    
    \subsection{Datasets and Evaluation Metrics}
    We evaluate GLINT-RU based on three benchmark datasets ML-1M ~\footnote{https://grouplens.org/datasets/movielens/}, Amazon-Beauty and Amazon video Games~\footnote{https://cseweb.ucsd.edu/~jmcauley/datasets.html\#amazon\_reviews}. Below is the basic introduction of MovieLens-1M, Amazon Beau-ty, and Amazon Video Games datasets.
\begin{itemize}[leftmargin=*]
    \item \textbf{MovieLens 1M:} comprises user ratings of movies. It includes about 1 million anonymous ratings from users who joined MovieLens. The dataset provides information about user IDs, movie IDs, ratings, and timestamps. It is widely used for research in collaborative filtering and recommendation systems.
    \item \textbf{Amazon Beauty:} is a subset of the Amazon product data, focusing specifically on beauty products. It includes user reviews and ratings of various beauty products available on Amazon. The dataset contains metadata such as product descriptions, categories, prices, and brand information. It is valuable for research in sentiment analysis, product recommendations, and consumer behavior analysis within the beauty industry.
    \item \textbf{Amazon Video Games:} is a subset of the  Amazon product data, focusing specifically on video game products. It includes user reviews and ratings of various video game products available on Amazon. The dataset contains metadata such as product descriptions, categories, price, and platform information. It is also valuable for research in sentiment analysis, product recommendation, and consumer behavior analysis in the video games domain.
\end{itemize}

    Statistical information of the three datasets is shown in Table \ref{tab:data}, where all the three datasets are sparse. The two Amazon datasets have relatively small data volumes, about 200,000, but they have a large number of users and items, and the average length of interaction sequences is short. In contrast, ML-1M has a larger data volume, reaching 1 million, but the number of users and items is small, with much longer interaction sequence lengths.

    We adopt Recall, Mean Reciprocal Rank (MRR), and Normalized Discounted Cumulative Gain (NDCG) as the evaluation metrics for our experiments. The interactions are grouped by users chronologically, and the datasets are split by the leave-one-out strategy~\cite{Linrec}. To be more specific, the penultimate item of the interaction sequence is used for validation. Therefore, the size of validation set is determined by the number of users. 
    
    \subsection{Baselines}
In this paper, we compare our GLINT-RU with two types of baseline models, i.e., traditional SRS models and efficient SRS models. The traditional SRS model includes various SRS benchmarks, while the efficient SRS models improve existing model's structure and computational methods, thus significantly enhancing the model efficiency. Adopted baselines are listed as follows:

\noindent\textbf{Traditonal SRS models}\\
(1) \textbf{GRU4Rec}~\cite{GRU4Rec}: utilizes GRUs to capture sequential dependencies within user interactions for session-based recommendations. \\
(2) \textbf{BERT4Rec} ~\cite{bert4rec}: adapts the Bidirectional Encoder Representations from Transformers (BERT) architecture to model user behaviors for personalized recommendation. \\
(3) \textbf{SASRec}~\cite{Kang01}: captures long-term and short-term user preferences by applying a multi-head attention mechanism.

\noindent\textbf{Efficient SRS Models} \\
(1) \textbf{LinRec}~\cite{Linrec}: reduces the computational costs substantially by changing the dot product of attention mechanism in the transformer-based models. We select SASRec as the backbone of LinRec. \\
(2) \textbf{SMLP4Rec} ~\cite{smlp4rec} uses a tri-directional fusion scheme to learn correlations on sequence, channel, and feature dimensions efficiently. \\
(3) \textbf{LightSAN} ~\cite{lightsan}  projects the initial interactions into representations with shorter lengths, which is also an efficient approach for transformer-based models.  \\
(4) \textbf{Mamba4Rec} ~\cite{mamba4rec}: explores the potential of selective SSMs for efficient sequential recommendation, which substantially improve the efficiency of SRS models.
    
    
\begin{table}[t]
    \caption{\textbf{Statistical Information of Adopted Datasets.}}
    \label{tab:data}
    \renewcommand{\arraystretch}{1}
    \resizebox{\linewidth}{!}{
    \begin{tabular}{cccccc}
        \toprule
        Datasets& \# Users & \# Items &\# Interactions & Avg.Length & Sparsity\\
        \midrule
        ML-1M & 6,041 & 3,707 & 1,000,209 & 165.60 & 95.53\%\\
        Beauty & 22,364 & 12,102 & 198,502 & 8.88 & 99.93\%\\
        Video Games& 24,304 & 10,673 & 231,780 & 9.54 & 99.91\%\\
        \bottomrule
    \end{tabular}}
    \vspace{-4mm}
    \end{table}
    \begin{table*}
		\caption{\textbf{Overall performance comparison between GLINT-RU and baselines.}}
		\label{tab:performance}
        \renewcommand{\arraystretch}{1.05}
        \resizebox{\linewidth}{!}{
		\begin{tabular}{cccccccccc}
			\toprule
			\multirow{2}{*}{Models} & \multicolumn{3}{c}{ML-1M} & \multicolumn{3}{c}{Amazon Beauty}&\multicolumn{3}{c}{Amazon Video Games}\\
            \cline{2-10}
            \vspace{-3mm}\\
            & Recall@10 & MRR@10 & NDCG@10 &Recall@10 & MRR@10 & NDCG@10 & Recall@10 & MRR@10 & NDCG@10\\
			\midrule
            GRU4Rec & 0.6954 & 0.4055 & 0.4748 & 0.3851 & 0.1891 & 0.2351 & 0.6028 & 0.2929 & 0.3660\\
            BERT4Rec & 0.7119 & 0.4041 & 0.4776 & 0.3478 & 0.1584 & 0.2027 & 0.5490 & 0.2541 & 0.2916\\
            SASRec & 0.7205 & 0.4251 & 0.4958 & 0.4332 & 0.2325 & 0.2798 & 0.6459 & 0.3404 & 0.4128\\
            LinRec & 0.7184 & 0.4316 & 0.5002 & 0.4270 & 0.2314 & 0.2775 & 0.6384 & 0.3355 & 0.4073\\
            LightSANs & 0.7195 & 0.4314 & 0.5003 & 0.4406 & 0.2358 & 0.2840 & \underline{0.6488} & 0.3415 & 0.4142\\
            SMLP4Rec & 0.6753 & 0.3870 & 0.4558 & \underline{0.4457} & \underline{0.2408} & \underline{0.2891} & 0.6480 & \underline{0.3484} & \underline{0.4195}\\
			Mamba4Rec & \underline{0.7238} & \underline{0.4368} & \underline{0.5054} & 0.4233 & 0.2213 & 0.2689 & \underline{0.6488} & 0.3389 & 0.4123\\
            GLINT-RU & \textbf{0.7379}$^{\ast}$ & \textbf{0.4517}$^{\ast}$ & \textbf{0.5202}$^{\ast}$ & \textbf{0.4472}$^{\ast}$ & \textbf{0.2498}$^{\ast}$ & \textbf{0.2964}$^{\ast}$ & \textbf{0.6573}$^{\ast}$ & \textbf{0.3549}$^{\ast}$ & \textbf{0.4266}$^{\ast}$\\
            \midrule
            Improv. & 1.95\% & 3.30\% & 2.93\% & 0.34\% & 3.74\% & 2.53\% & 1.31\% & 1.87\% & 1.69\%\\
			\bottomrule
		\end{tabular}}
        \begin{tablenotes}
            \small\centering
            \item[*] Recommendation performance of GLINT-RU and existing state-of-the-art benchmark SRS models have been shown. $``\ast"$ indicates the improvements are \textbf{statistically significant} (i.e., two-sided t-test with $p$ < 0.05) over baselines). The best results are bolded, and the second-best are underlined.
        \end{tablenotes}
	\end{table*}
    \subsection{Implementation}
    In this subsection, we introduce the implementation details of the GLINT-RU. We use Adam optimizer ~\cite{Adam} with the learning rate of 0.001 for our training process. Both the train and evaluation batch size are set as 2048. The hidden size is set as 128 for ML-1M and 64 for Amazon Beauty and Video Games. As is shown in Table \ref{tab:data}, the average length of ML-1M, Beauty, and Video Games are 165, 8.88, and 9.54, so we set the maximum sequence length as 200 for ML-1M, and 100 for the two amazon datasets. We adopt the dropout rate of 0.5 for Amazon datasets considering their high level of sparsity, compared with 0.2 for ML-1M. We construct the attention-based baselines with 2 transformer layers so that they can achieve high performance. Other implementation details follow the settings of original papers~\cite{mamba4rec, smlp4rec, Linrec}. Please find the complete hyperparameter settings and more implementation details in \textbf{Appendix \ref{hypersetting}}.
    
    \subsection{Overall Performance (RQ1)}
    In this subsection, we compare the performance of GLINT-RU with both traditional recommendation frameworks and state-of-the-art efficient models. The results, as shown in Table ~\ref{tab:performance}, demonstrate the effectiveness of GLINT-RU on metrics Recall@10, MRR@10, and NDCG@10 in bold.
    According to the above table, it is evident that GLINT-RU defeats all the selected transformer-, RNN-, MLP-, and SSM-based baselines. We improve the performance upper bound of efficient recommendation models by $0.34\%\sim 3.74\%$. 
    
    Traditional RNN-based models, like GRU4Rec, might have difficulty in dealing with complex user behaviors. Although it can capture the long-term dependencies from long sequences, it struggles to learn effectively from extremely sparse datasets. Traditional attention-based methods like BERT4Rec and SASRec have great performance on the three datasets, but it is quite inefficient due to the high computational complexity of the attention mechanism. 
    
    Efficient model LinRec changes the softmax operation, and takes attention scores from more positions into consideration, improving the performance on long-term sequential recommendation tasks compared with its backbone SASRec. SMLP4Rec and Mamba4Rec achieve impressive performance on the three datasets. However, Mamba4Rec shows enhanced proficiency in modeling long sequences (ML-1M) but exhibits low performance in relatively short sequences (Beauty and Video Games). Conversely, the SMLP4Rec shows superior performance in tasks with short sequences while being less effective with longer sequences. Additionally, SMLP4Rec requires features from the items to enhance its performance. Uniquely, GLINT-RU integrates the advantages of a linear attention mechanism and dense selective GRU module, adaptively extracting dependencies from recurrent latent item representations, fine-grained positional representations, and important semantic features. The gate mechanism employed in GLINT-RU substantially enhances its ability to filter the information based on the dynamic data environment and mix the experts according to the data adaptively.

    In summary, GLINT-RU as a novel efficient framework, shows its superiority over state-of-the-art baselines. 
    This underscores the potential of dense selective GRU and models with hybrid modules as more powerful tools for recommendation tasks.

    \begin{table}
		\caption{\textbf{Efficiency comparison.}}
		\label{tab:efficiency}
        \renewcommand{\arraystretch}{1}
        \resizebox{\linewidth}{!}{
		\begin{tabular}{ccccc}
			\toprule
			Datasets & Model & Infer.& Training & GPU Memory\\
			\midrule
			\multirow{8}{*}{ML-1M} & BERT4Rec & 88ms & 285s/epoch & 21.51GB\\
            & SASRec & 55ms & 172s/epoch & 21.51GB\\
            & LinRec & \underline{37ms} & \underline{101s/epoch} & 11.67GB\\
            & LightSANs & 43ms & 130s/epoch & 16.99GB\\
            & SMLP4Rec & 51ms & 151s/epoch & 16.13GB\\
            & Mamba4Rec & 41ms & 108s/epoch & \textbf{7.72G}\\
            & GLINT-RU & \textbf{31ms} & \textbf{86s/epoch} & \underline{8.81G}\\
            \cline{2-5}
            \vspace{-3mm}\\
            & Improv. & 16.22\% & 17.44\% & - \\
            \midrule
			\multirow{8}{*}{Beauty} & BERT4Rec & 1372ms & 13s/epoch & 11.69GB\\
            & SASRec & 444ms & 8.1s/epoch & 7.67GB\\
            & LinRec & \underline{340ms} & 5.6s/epoch & 4.14GB\\
            & LightSANs & 427ms & 7.2s/epoch & 4.57GB \\
            & SMLP4Rec & 361ms & 5.3s/epoch & 2.95GB\\
            & Mamba4Rec & 351ms & \underline{4.5s/epoch} & \textbf{2.32G}\\
            & GLINT-RU & \textbf{278ms} & \textbf{3.8s/epoch} & \underline{2.62G}\\
            \cline{2-5}
            \vspace{-3mm}\\
            & Improv. & 18.24\% & 15.56\% & - \\
            \midrule
			\multirow{8}{*}{Video Games} & BERT4Rec & 1290ms & 15s/epoch & 10.98GB\\
            & SASRec & 406ms & 9.6s/epoch & 7.65GB\\
            & LinRec & 327ms & 6.6s/epoch & 4.13GB\\
            & LightSANs & 369ms & 8.3s/epoch & 4.55GB \\
            & SMLP4Rec & 389ms & 9.1s/epoch & 3.38GB\\
            & Mamba4Rec & \underline{309ms} & \underline{5.6s/epoch} & \textbf{2.28G}\\
            & GLINT-RU & \textbf{247ms} & \textbf{4.5s/epoch} & \underline{2.49G}\\
            \cline{2-5}
            \vspace{-3mm}\\
            & Improv. & 20.06\% & 19.64\% & - \\
			\bottomrule
		\end{tabular}}
        \begin{tablenotes}
            \small\centering
            \item[*] 
            Inference time of each mini-batch (batch size = 2048), training time and GPU memory of GLINT-RU and other baseline models. The best results are bolded, and the second best results are underlined.
        \end{tablenotes}
        \vspace{-5mm}
	\end{table}
    \subsection{Efficiency Comparison (RQ2)}
    In this subsection, we analyze the efficiency of GLINT-RU and state-of-the-art sequential recommendation models. We evaluate the model efficiency according to the inference time of each mini-batch, training time, and GPU memory occupation.
    
    The results, shown in Table~\ref{tab:efficiency}, provide several valuable insights: Firstly, by utilizing the efficient Dense GRU module and linear attention module, GLINT-RU dramatically reduces the training time and inference time, improving the training inference time by $15\%\sim 20\%$ compared with most efficient recommendation baseline models. In addition, due to the low computational cost of GRU and linear attention mechanism, GLINT-RU exhibits minimal GPU memory consumption, which is comparable to the state-of-the-art SSM-based efficient model Mamba4Rec. In Section \ref{complexity}, we demonstrate that the GLINT-RU exhibits low theoretical computational complexity, as we employ the parallel networks and efficient GRU as core components of our recommender system. The theoretical analysis has been verified by the results in Table \ref{tab:efficiency}.

    Traditional transformer-based recommendation models like SASRec and BERT4Rec suffer from extended inference time and high GPU memory occupation. When processing long sequential data, the conventional attention mechanism falls behind novel efficient models due to its high computational cost. Among all the baseline models, the SSM-based Mamba4Rec framework exhibits impressive efficiency, but Mamba requires complex mathematical computation, which slows down its inference and training speed. Additionally, LinRec suffers from the inherent shortage of its backbone SASRec that requires stacked transformer layers to enhance the model performance. Such large transformer structures extend both the inference and training time. Although SMLP4Rec achieves high performance in the model accuracy, it struggles to train and inference efficiently, especially when processing long-term sequential data.

    \begin{table}
		\caption{\textbf{Ablation study for Components of GLINT-RU.}}
		\label{tab:ablation}
        \renewcommand{\arraystretch}{1}
        \resizebox{\linewidth}{!}{
		\begin{tabular}{cccc}
			\toprule
			Model Components & Recall@10 & MRR@10 & NDCG@10\\
			\midrule
			Default & \textbf{0.7379$^{\ast}$} & \textbf{0.4517$^{\ast}$} & \textbf{0.5202$^{\ast}$}\\
            w/o Gated MLP (Light GLINT-RU) & \underline{0.7260} & \underline{0.4369} & \underline{0.5060}\\
            w/o Attention & 0.7195 & 0.4312 & 0.5001\\
            w/o GRU & 0.6762 & 0.3913 & 0.4593\\
            w/o Temporal Conv1d & 0.7232 & 0.4322 & 0.5019\\
			\bottomrule
		\end{tabular}}
        \begin{tablenotes}
            \small\centering
            \item[*] $``\ast"$ indicates the improvements are \textbf{statistically significant} (i.e., two-sided t-test with $p$ < 0.05) over baselines)
        \end{tablenotes}
	\end{table}
    \subsection{Ablation Study (RQ3)}
    In the ablation study, we remove the gated MLP block, linear attention expert, dense selective GRU expert, and the two temporal convolution layers individually to verify the efficacy of each component. We conduct the ablation study on the ML-1M dataset, and the results are outlined in Table~\ref{tab:ablation}, providing insightful observations.

    The results verify the essential role of the dense selective GRU module, as the performance of the model will dramatically decrease without the GRU module. This reveals the insights that the gated GRU module effectively captures the dependencies of interactions with fine-grained positional representations. The linear attention mechanism understands the interactions of relevant items in the sequence. As is shown in Table \ref{tab:ablation}, it improves the performance of GLINT-RU to some extent. Adding a temporal convolution layer incorporates context information from adjacent items, resulting in an enhancement in model performance. In addition, the gated MLP block plays a similar role as the feed-forward network in our framework, which filters complex information from the expert mixing block. It is noteworthy that even without the gated MLP block our framework still outperforms all the state-of-the art efficient models, demonstrating its inherent remarkable superiority for sequential recommendation tasks. After we remove the gated MLP, the GPU memory occupation of GLINT-RU becomes \textbf{7.63GB}, less than Mamba4Rec shown in Table \ref{tab:efficiency}, and the inference time will be reduced to \textbf{241ms}. We name this framework ``Light GLINT-RU'', which is more applicable to resource-constrained scenarios. 
We further discuss the ablation study on activation functions in \textbf{Appendix~\ref{sec:Activation}}, where we highlight their impact on performance. Additionally, we provide a detailed analysis of the ablation study conducted on the Amazon Beauty and Amazon Video Games datasets in \textbf{Appendix~\ref{sec:AblationStudyAmazon}}, emphasizing the contributions of each component to the overall performance of GLINT-RU.
    
    \subsection{Parameter Analysis}
    We conduct parameter analysis on the dataset Amazon Beauty. We will first analyze the impact of the crucial hyperparameter kernel size $k$ in GLINT-RU,
    and then we will analyze the model performance as the hidden size $d$ and number of GLINT-RU layers $L$ changes. The discussion on these parameters provides valuable insights into the superiority of GLINT-RU. 

\begin{figure}[t]
        \centering
        \includegraphics[width=\linewidth]{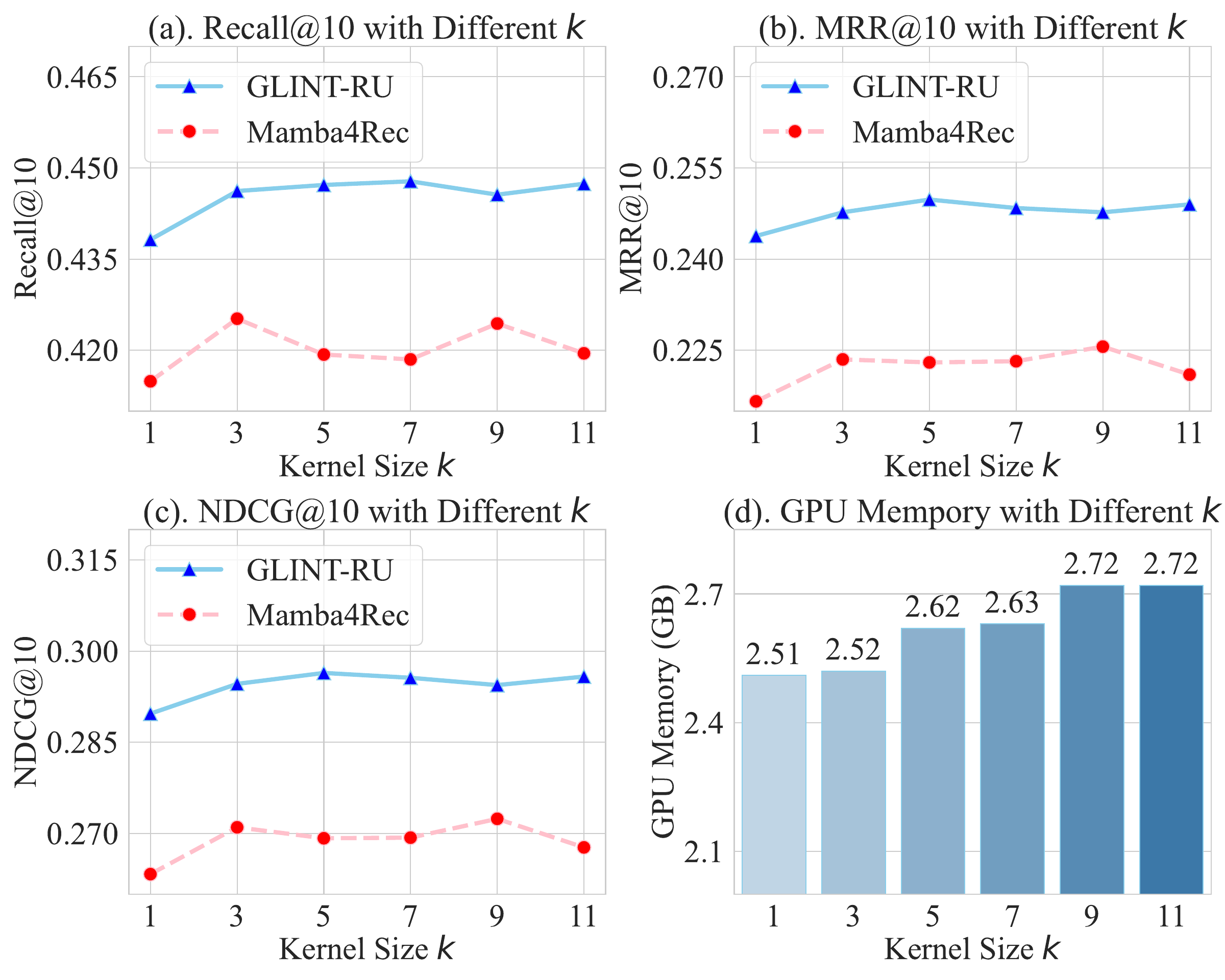}
        \caption{\textbf{Impacts of kernel size $k$ of the temporal convolution on the performance of GLINT-RU and Mamba4Rec and the GPU Memory occupation of GLINT-RU.}}
    \label{fig:parameter1}
    \vspace{-5mm}
    \end{figure}

\begin{figure}[t]
        \centering
        \includegraphics[width=\linewidth]{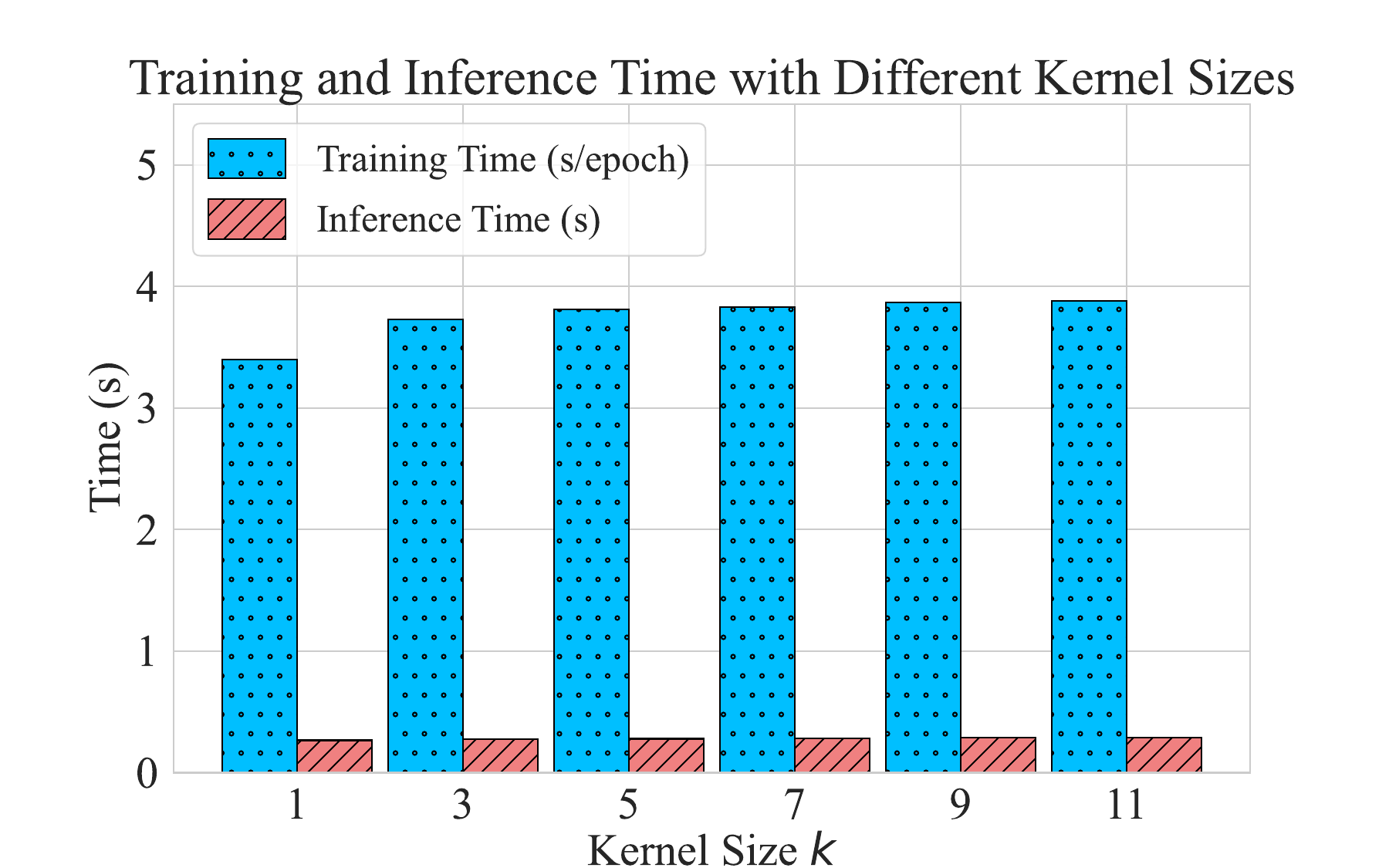}
        \caption{\textbf{Impacts of kernel size $k$ of the temporal convolution on the training and inference time.}}
    \label{fig:parameter2}
    \vspace{-5mm}
    \end{figure}

    \noindent \textbf{Kernel size $k$}.
    The impacts of the parameter $k$ on the model performance are shown in Figure \ref{fig:parameter1} and \ref{fig:parameter2}.
    \begin{figure*}[t]
        \centering
        \includegraphics[width=\linewidth]{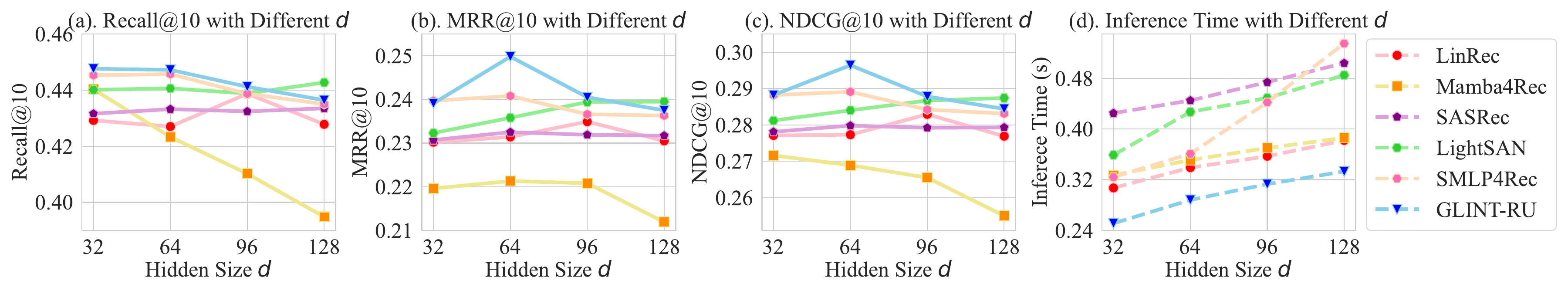}
        \caption{\textbf{Impacts of hidden size $d$ on the performance of GLINT-RU and state-of-the-art baselines.}}
    \label{fig:parameter_d}
    \vspace{-2mm}
    \end{figure*}
    Figure \ref{fig:parameter1}.(a)-(c) displays the model performance of GLINT-RU with different kernel sizes. Our model exhibits stable and high performance, providing a wide range of kernel sizes. This finding indicates the robustness of the GLINT-RU framework. This enhancement in performance can be attributed to the fact that a larger kernel size aggregates information from more items, thereby learning a more extensive context into the hidden state. However, as the kernel size continues to expand, dense selective GRU might incorporate irrelevant data into the output state, which might lead to a marginal decline in the accuracy. Mamba4Rec as a novel efficient SSM-based model, also employs a temporal convolution layer in the model structure. As the kernel size changes, the performance of Mamba4Rec becomes quite unstable compared with our GLINT-RU. Our GLINT-RU consistently outperforms Mamba4Rec across all kernel sizes, further demonstrating the superiority of this novel GLINT-RU framework.
    
    As can be seen in the hist plots in Figure \ref{fig:parameter1}.(d) and Figure \ref{fig:parameter2}, increasing the kernel size has slight impacts on the training/inference time of each mini-batch and the GPU memory occupation, which further verifies the efficiency and stability of our model. 

    \noindent\textbf{Hidden size $d$.}
    We change the model size by using different hidden sizes $d$ and compare the performance of GLINT-RU with state-of-the-art baselines. The results shown in Figure \ref{fig:parameter_d}.(a)-(c) indicate that GLINT-RU achieves state-of-the-art performance at small hidden sizes. When $d$ is set as 64, the predictive efficacy of the GLINT-RU substantially surpasses the upper bounds of performance achieved by other baseline models. Achieving excellent results with a relatively small model dimension illustrates the superior expressiveness and significant advantages of the GLINT-RU. SMLP4Rec requires additional features to enhance the model accuracy, and its inference speed is significantly affected by the variation in the hidden size, demonstrating relatively low efficiency. The prediction accuracy of Mamba4Rec is inferior and decreases dramatically as the hidden size $d$ gets larger, indicating that it struggles to learn effective information from short behavior sequences. Attention-based models SASRec, LightSAN, and LinRec show more stable results, but they can only capture item interactions to predict the ratings and require stacked transformers to achieve relatively high performance, which has negative impacts on model efficiency in Figure \ref{fig:parameter_d}.(d).

    \noindent\textbf{Number of Layers $L$.} 
    We increase the number of GLINT-RU layers from 1 to 4 and observe the model performance and efficiency. The results and the experimental details are illustrated in \textbf{Appendix \ref{layernum}}. The results indicate that our single-layer GLINT-RU can achieve high model efficiency and accuracy simultaneously.
    
In summary, GLINT-RU effectively combines the long/short-term dependencies with effective positional representations and important interactions which enables it to be an accurate and efficient SRS under a broad range of parameter choices.

\section{Related Works}
\noindent\textbf{Traditional Sequential Recommendation Models}
Transformers and RNNs are widely recognized as popular and effective frameworks for sequential recommendation tasks. Numerous studies have been conducted to explore the potential of transformers in enhancing recommendation performance. The transformer-based SASRec~\cite{Kang01} predicts user preferences by leveraging multi-head attention to model both long- and short-term interactions. SSE-PT~\cite{sse-pt} further improves recommendation accuracy by incorporating user embeddings into the transformer architecture. Similarly, MB-STR~\cite{mb-str}, a transformer-based variant, captures diverse user behavior dynamics and effectively mitigates the challenges posed by data sparsity.  
On the other hand, RNN-based methods like GRU4Rec~\cite{GRU4Rec} model dependencies among items through gated recurrent units, enabling them to learn sequential patterns. 
HRNN~\cite{HRNN}, an advanced RNN-based framework, integrates an additional GRU layer to extract session-level information and dynamically track user preferences over time.  
In comparison to these models, GLINT-RU demonstrates superior capability in capturing high-quality semantic features and fine-grained positional representations, leading to improved model accuracy and performance.

\noindent\textbf{Efficient Recommendation Models}
To tackle the high computational complexity of existing SRSs, and accelerate the inference speed for real-world applications, researchers endeavor to invent increasingly efficient models. AutoSeqRec ~\cite{autoseqrec} is constructed based on Autoencoder, which is an innovative work that captures long-term preferences through collaborative filtering. To further improve the model efficiency, DMAN~\cite{DMAN} combines the long-term attention net and recurrent attention net to memorize users' interests dynamically and support efficient inference. LinRec ~\cite{Linrec} cuts down the computational cost of transformer-based backbones by changing the dot product in the attention mechanism. 
SSM-based models like Mamba4Rec~\cite{mamba4rec} and RecMamba~\cite{MambaRec} utilize selective state space models to achieve high performances and high efficiency, becoming emerging powerful tools for sequential recommendation tasks. FMLP-Rec~\cite{FMLP} with learnable filters and SMLP4Rec~\cite{smlp4rec} with diverse mixers are representative pure MLP-based efficient SRSs. LRURec~\cite{LRU} is constructed based on linear recurrent units and achieves fast inference by employing recursive parallelization. 
These models often require stacked network structures to achieve better results, and their performance is not stable enough when faced with different sequence lengths. However, GLINT-RU only needs one layer to achieve stable high performance.

\section{Conclusion}
In this paper, we have presented an innovative dense selective GRU framework, GLINT-RU for sequential recommendation tasks. Due to the paralleled network design and implementation of efficient dense selective GRU with linear complexity, the computational cost can be substantially reduced, resulting in state-of-the-art inference speed. Additionally, our GLINT-RU models improve the quality of semantic features and fine-grained positional information for recommendation tasks. Gate mechanisms are widely applied in GLINT-RU, deeply filtering and selecting information. It uses a dense selective GRU that aggregates information from adjacent items and generates high-quality latent item representations based on dependencies with fine-grained positional information to learn context information. By mixing dense selective GRU with linear attention, GLINT-RU can capture important interactions and item dependencies simultaneously. Our extensive experiments demonstrate that GLINT-RU achieves outstanding performance, not only improving model accuracy but also accelerating training and inference speed dramatically. These results underscore GLINT-RU's potential to become a novel, stable, and efficient framework in various scenarios. As an efficient recommender system that defeats state-of-the-art models, we believe our novel framework will become a valuable foundation.
\section*{Acknoledgement}
This research was partially supported by Research Impact Fund (No.R1015-23), APRC - CityU New Research Initiatives (No.9610565, Start-up Grant for New Faculty of CityU), CityU - HKIDS Early Career Research Grant (No.9360163), Hong Kong ITC Innovation and Technology Fund Midstream Research Programme for Universities Project (No.ITS/034/22MS), Hong Kong Environmental and Conservation Fund (No. 88/2022), and SIRG - CityU Strategic Interdisciplinary Research Grant (No.7020046), Collaborative Research Fund (No.C1043-24GF), Huawei (Huawei Innovation Research Program), Tencent (CCF-Tencent Open Fund, Tencent Rhino-Bird Focused Research Program), Ant Group (CCF-Ant Research Fund, Ant Group Research Fund), Alibaba (CCF-Alimama Tech Kangaroo Fund No. 2024002), CCF-BaiChuan-Ebtech Foundation Model Fund, and Kuaishou.
\balance
\bibliographystyle{ACM-Reference-Format}
{\bibliography{sample-base}}
\newpage
\begin{appendices}
\begin{table*}
	\caption{\textbf{Hyperparameter settings of GLINT-RU and the baselines.}}
	\label{hyper}
    \renewcommand{\arraystretch}{1}
	\begin{tabular}{cccccccccc}
	\toprule
	Model & Dataset &  hidden size &  weight decay & dropout & layers & heads & max length & kernel size & train\&eval batch size \\
	\midrule
	\multirow{3}{*}{GRU4Rec} & ML-1M & 128 & 0 & 0.2 & 2 & - & 200 & - & [2048, 2048]\\
    & Beauty & 64 & 0 & 0.5 & 2 & - & 100 & - & [2048, 2048]\\
    & Games & 64 & 0 & 0.5 & 2 & - & 100 & - & [2048, 2048]\\
    \multirow{3}{*}{BERT4Rec} & ML-1M & 128 & 0 & 0.2 & 2 & 8 & 200 & - & [2048, 2048]\\
    & Beauty & 64 & 0 & 0.5 & 2 & 8 & 100 & - & [2048, 2048]\\
    & Games & 64 & 0 & 0.5 & 2 & 8 & 100 & - & [2048, 2048]\\
    \multirow{3}{*}{SASRec} & ML-1M & 128 & 0 & 0.2 & 2 & 8 & 200 & - & [2048, 2048]\\
    & Beauty & 64 & 0 & 0.5 & 2 & 8 & 100 & - & [2048, 2048]\\
    & Games & 64 & 0 & 0.5 & 2 & 8 & 100 & - & [2048, 2048]\\
    \multirow{3}{*}{LinRec} & ML-1M & 128 & 0 & 0.2 & 2 & 8 & 200 & - & [2048, 2048]\\
    & Beauty & 64 & 0 & 0.5 & 2 & 8 & 100 & - & [2048, 2048]\\
    & Games & 64 & 0 & 0.5 & 2 & 8 & 100 & - & [2048, 2048]\\
    \multirow{3}{*}{LightSANs} & ML-1M & 128 & 0 & 0.2 & 2 & 8 & 200 & - & [2048, 2048]\\
    & Beauty & 64 & 0 & 0.5 & 2 & 8 & 100 & - & [2048, 2048]\\
    & Games & 64 & 0 & 0.5 & 2 & 8 & 100 & - & [2048, 2048]\\
    \multirow{3}{*}{SMLP4Rec} & ML-1M & 128 & 0 & 0.2 & 2 & - & 200 & - & [2048, 2048]\\
    & Beauty & 64 & 0 & 0.5 & 2 & - & 100 & - & [2048, 2048]\\
    & Games & 64 & 0 & 0.5 & 2 & - & 100 & - & [2048, 2048]\\
    \multirow{3}{*}{Mamba4Rec} & ML-1M & 128 & 0 & 0.2 & 2 & - & 200 & 3 & [2048, 2048]\\
    & Beauty & 64 & 0 & 0.5 & 2 & - & 100 & 3 & [2048, 2048]\\
    & Games & 64 & 0 & 0.5 & 2 & - & 100 & 3 & [2048, 2048]\\
    \multirow{3}{*}{GLINT-RU} & ML-1M & 128 & 0 & 0.2 & 2 & 8 & 200 & 3 & [2048, 2048]\\
    & Beauty & 64 & 0 & 0.5 & 2 & 8 & 100 & 3 & [2048, 2048]\\
    & Games & 64 & 0 & 0.5 & 2 & 8 & 100 & 3 & [2048, 2048]\\
    \bottomrule
	\end{tabular}
\end{table*}
\section{Model Training Methods}
\label{training}
In this section, we introduce the details of the training methods of GLINT-RU, including the loss function and the optimization algorithm of GLINT-RU displayed through pseudo-code. 

Based on the prediction score in Eq.(\ref{score}), we adopt the cross entropy loss function defined below for training GLINT-RU:
\begin{equation}
\label{loss}
    \mathcal{L} = \sum\limits_{i=1}^{n_i} \left( y\log(\hat{y}_i) + (1-y) \log (1-\hat{y}_i) \right),
\end{equation}
where $y$ are the ground-truth labels of the user-item interactions. Now we can train GLINT-RU by following the guidance in Algorithm \ref{alg:algorithm}. As can be seen in the optimization algorithm, the training process of GLINT-RU can be easily implemented. The paralleled expert linear attention and dense selective GRU can be balanced by updating mixing weights. Additionally, to help GLINT-RU find a better local optimal solution, we follow the suggestion of ~\cite{Xavier} to initialize the model parameters $\boldsymbol{\mathcal{W}}$,  which helps prevents the vanishing/exploding gradient problem.

\begin{algorithm}[t]
    \caption{Optimization Algorithm of GLINT-RU}
    \begin{flushleft}
        \textbf{Input:} User-item interactions $\boldsymbol{\mathcal{S}}=[s_1,\cdots, s_{\vert \mathcal{U}\vert }]$, corresponding ground-truth labels $y$\\
        \textbf{Output:} Well-trained weights $\boldsymbol{\mathcal{W}}^{\ast}$ of model $f$ 
    \end{flushleft}
    \begin{algorithmic}[1]
        \State Randomly initialize the model parameters $\boldsymbol{\mathcal{W}}^{\ast}$.
        \State $t$ = 0 ($t$ represents the number of iteration)
        \For{Epoch in $1,\dots,$ max epoch}
            \For{Batch in $1,\dots,$ batch number}
            \State Sample training batch data $\boldsymbol{S}^{\prime}$ from $\boldsymbol{\mathcal{S}}$.
            \State Generate predictions from $f(\boldsymbol{S}^{\prime})$.
            \State Calculate Loss according to Eq.(\ref{loss}).
            \State Update the expert mixing weights $\alpha_i$ using Eq.(\ref{mixing}).
            \State Update parameters $\boldsymbol{\mathcal{W}}$ via minimizing the loss $\mathcal{L}$.\
            \If{Converged}
                \State Return $f$ and parameters $\boldsymbol{\mathcal{W}}^{\ast}$.
            \EndIf
            \EndFor
        \EndFor
        \State Return $f$ and parameters $\boldsymbol{\mathcal{W}}^{\ast}$.
    \end{algorithmic}
    \label{alg:algorithm}
\end{algorithm}

\section{Hyperparameter Settings}
\label{hypersetting}
In this section, we will introduce the hyperparameter settings of our GLINT-RU and the baselines, and offer implementation details for reproducibility.

The parameter settings of all baselines are displayed in Table \ref{hyper}. In addition to the hyperparameter settings in Table \ref{hyper}, we also set the number of interests in LightSANs to 5, as the the average sequence length is less than 10, which is shown in Table \ref{tab:data}. We set the expansion factor in Mamba4Rec as 2, and set the expansion factor in SMLP4Rec as 1 to avoid the GPU memory occupation error. The feature selection in SMLP4Rec is consistent with the settings in the original paper.~\cite{smlp4rec}. For ML-1M dataset, we choose ['genre', 'movie title', 'release year'] as the input features in the SMLP structure. And for Amazon Beauty and Games, we select ['categories', 'brand'] as the input features. These features are employed to enhance the prediction performance from the item ids in SMLP4Rec.

\begin{table}[t]
    \centering
    \caption{Ablation study on activation functions in the dense selective gate.}
    \label{tab:activation_ablation}
    \begin{tabular}{lccc}
        \hline
        \textbf{Method} & \textbf{Recall@10} & \textbf{MRR@10} & \textbf{NDCG@10} \\
        \hline
        Default (SiLU)  & 0.4472 & 0.2498 & 0.2964 \\
        ReLU            & 0.4293 & 0.2300 & 0.2770 \\
        Sigmoid         & 0.4328 & 0.2319 & 0.2793 \\
        ELU             & 0.4308 & 0.2321 & 0.2789 \\
        \hline
    \end{tabular}
\end{table}

\section{Ablation Study: Activation Functions}
\label{sec:Activation}
Unlike ReLU, which has a sharp kink at zero, SiLU is smooth and continuously differentiable. This smoothness facilitates better gradient flow during backpropagation, leading to faster and more stable training. Additionally, SiLU mitigates the vanishing gradient problem more effectively than Sigmoid and Tanh, as its gradient does not saturate as quickly, allowing for more efficient training of deeper networks. Incorporating SiLU as the activation function in the gate introduces non-linearity while retaining important item information and filtering out irrelevant item information, which is advantageous for complex tasks.
Studies have demonstrated that SiLU enhances training stability and performance in deep learning models. In practical applications, user-item interactions often contain noise, as some selected items may be irrelevant. Under such conditions, employing SiLU typically achieves higher accuracy compared to traditional activation functions like ReLU and ELU.
To further verify the effectiveness of the SiLU activation function, we conducted additional experiments where SiLU was replaced by ELU, Sigmoid, and ReLU in the dense selective gate. The results are presented in Table~\ref{tab:activation_ablation}.

\section{Parameter Analysis: Layer Number}
\label{layernum}
\begin{figure}[t]
        \centering
        \includegraphics[width=0.85\linewidth]{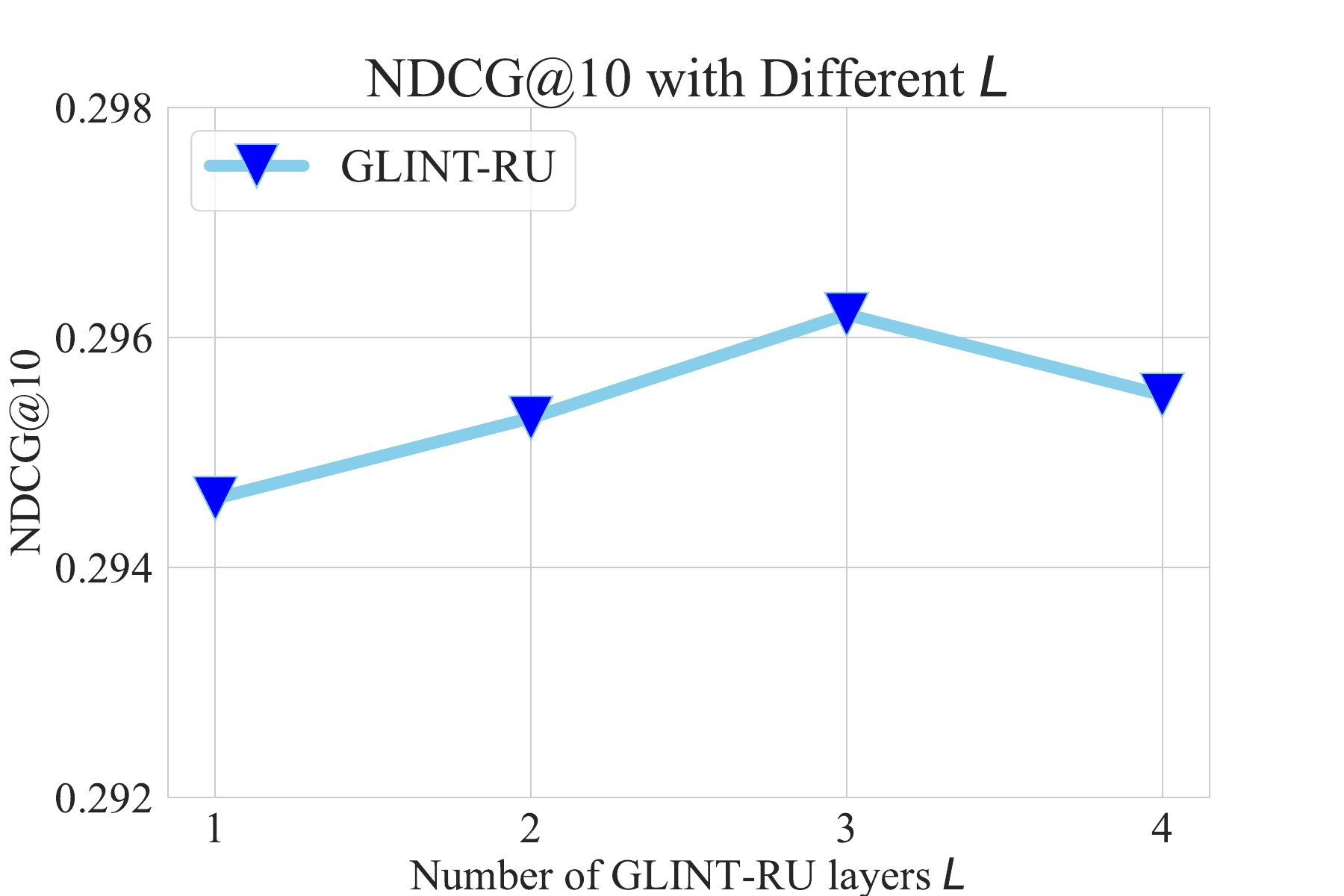}
        \caption{\textbf{Impacts of $L$ on the model performance.}}
    \label{fig:L}
\end{figure}

In this section, we conduct the parameter analysis on the number of layers $L$ to observe the model's accuracy and efficiency in stacked GLINT-RU. We change the number of GLINT-RU layers from 1 to 4, and evaluate the model performance with NDCG@10 using the dataset Amazon Beauty. The results are shown in Figure \ref{fig:L}.

From Figure \ref{fig:L}, it is evident that as the number of layers increases, the accuracy of the GLINT-RU model exhibits a relatively gradual upward trend. This indicates that stacking GLINT-RU layers contributes positively to improving accuracy. However, this improvement is not pronounced, and the model's performance may even slightly decline when the number of layers becomes excessively large. Additionally, we analyze the model efficiency of GLINT-RU in Table \ref{tab:efficiency_L}:
\begin{table}[t]
    \caption{\textbf{Efficiency analysis on the parameter $L$}}
    \label{tab:efficiency_L}
    \renewcommand{\arraystretch}{0.8}
    \resizebox{\linewidth}{!}{
    \begin{tabular}{cccc}
        \toprule
        $L$ & Inference Time & Training Speed & GPU Memory\\
        \midrule
        1 & 278ms & 3.8s/epoch & 2.62GB\\
        2 & 397ms & 6.2s/epoch & 5.17GB\\
        3 & 501ms & 8.9s/epoch & 6.97GB\\
        4 & 610ms & 11.8s/epoch & 8.86GB\\
        \bottomrule
    \end{tabular}}
    \vspace{-4mm}
\end{table}
The results in Table \ref{tab:efficiency_L} indicate that as the number of layers increases, the efficiency of GLINT-RU decreases rapidly. Combined with the previous model performance analysis, we can conclude that under resource constraints, using only one layer of efficient GLINT-RU can achieve high model efficiency and accuracy simultaneously.

\section{Ablation Study: Amazon Beauty Dataset and Video Games Dataset}
\label{sec:AblationStudyAmazon}
To further verify the effectiveness of the components in GLINT-RU, we conducted more ablation studies on the Amazon Beauty and Amazon Video Games datasets. The results are shown in Tables~\ref{tab:ablation_beauty} and~\ref{tab:ablation_videogames}. The results demonstrate that the gated GRU module plays a critical role in capturing dependencies among interactions and fine-grained positional representations. The linear attention mechanism enhances the model's ability to capture interactions between relevant items within the sequence. Adding a temporal convolution layer improves performance by incorporating contextual information from adjacent items. Finally, the gated MLP block filters complex information from the expert mixing block, further contributing to the overall performance.  
These findings indicate that all components of GLINT-RU are effective and collectively contribute to its high performance on both the Amazon Beauty and Amazon Video Games datasets.

\begin{table}[h!]
    \centering
    \caption{Ablation study results on Amazon Video Games.}
    \label{tab:ablation_videogames}
    \begin{tabular}{lccc}
        \hline
        \textbf{Method}        & \textbf{Recall@10} & \textbf{MRR@10} & \textbf{NDCG@10} \\
        \hline
        Default (GLINT-RU)     & 0.6573             & 0.3549          & 0.4266           \\
        w/o GRU                & 0.6379             & 0.3125          & 0.4023           \\
        w/o Linear Attention   & 0.6459             & 0.3375          & 0.4106           \\
        w/o Temporal Conv1d    & 0.6443             & 0.3351          & 0.4084           \\
        w/o Gated MLP          & 0.6416             & 0.3339          & 0.4068           \\
        \hline
    \end{tabular}
\end{table}

\begin{table}[h!]
    \centering
    \caption{Ablation study results on Amazon Beauty.}
    \label{tab:ablation_beauty}
    \begin{tabular}{lccc}
        \hline
        \textbf{Method}        & \textbf{Recall@10} & \textbf{MRR@10} & \textbf{NDCG@10} \\
        \hline
        Default (GLINT-RU)     & 0.4472             & 0.2498          & 0.2964           \\
        w/o GRU                & 0.4239             & 0.2298          & 0.2755           \\
        w/o Linear Attention   & 0.4296             & 0.2314          & 0.2781           \\
        w/o Temporal Conv1d    & 0.4246             & 0.2290          & 0.2751           \\
        w/o Gated MLP          & 0.4265             & 0.2308          & 0.2768           \\
        \hline
    \end{tabular}
\end{table}

\end{appendices}
\end{document}